\documentclass[%
 reprint,
superscriptaddress,
 amsmath,amssymb,
 aps,
pra,
]{revtex4-2}

\usepackage{graphicx}
\usepackage{float}
\usepackage{graphicx}
\usepackage{dcolumn}
\usepackage{bm}
\usepackage{verbatim}
\usepackage[dvipsnames]{xcolor}
\usepackage{hyperref}
\usepackage[mathlines]{lineno}
\usepackage{makecell}
\usepackage{ulem}
\usepackage{comment}
\begin{document}

\preprint{APS/123-QED}

\title{Energy-time Entanglement Coexisting with Fiber Optical Communication at Telecom C-band}

\author{Yun-Ru Fan}
\affiliation{Institute of Fundamental and Frontier Sciences, University of Electronic Science and Technology of China, Chengdu 610054, China}
\author{Yue Luo}
\affiliation{Institute of Fundamental and Frontier Sciences, University of Electronic Science and Technology of China, Chengdu 610054, China}
\author{Zi-Chang Zhang}
\affiliation{Institute of Fundamental and Frontier Sciences, University of Electronic Science and Technology of China, Chengdu 610054, China}
\author{Yun-Bo Li}
\affiliation{\mbox{Department of Fundamental Network Technology, China Mobile Research Institute, Beijing 100053, China.}}
\author{Sheng Liu}
\affiliation{\mbox{Department of Fundamental Network Technology, China Mobile Research Institute, Beijing 100053, China.}}
\author{Dong Wang}
\affiliation{\mbox{Department of Fundamental Network Technology, China Mobile Research Institute, Beijing 100053, China.}}
\author{Dechao Zhang}
\affiliation{\mbox{Department of Fundamental Network Technology, China Mobile Research Institute, Beijing 100053, China.}}
\author{Guang-Wei Deng}
\affiliation{Institute of Fundamental and Frontier Sciences, University of Electronic Science and Technology of China, Chengdu 610054, China}
\author{You Wang}
\affiliation{Institute of Fundamental and Frontier Sciences, University of Electronic Science and Technology of China, Chengdu 610054, China}
\affiliation{Southwest Institute of Technical Physics, Chengdu 610041, China}
\author{Hai-Zhi Song}
\affiliation{Institute of Fundamental and Frontier Sciences, University of Electronic Science and Technology of China, Chengdu 610054, China}
\affiliation{Southwest Institute of Technical Physics, Chengdu 610041, China}
\author{Zhen Wang}
\affiliation{\mbox{Shanghai Institute of Microsystem and Information Technology, Chinese Academy of Sciences, Shanghai 200050, China}}
\author{Li-Xing You}
\affiliation{\mbox{Shanghai Institute of Microsystem and Information Technology, Chinese Academy of Sciences, Shanghai 200050, China}}
\author{Chen-Zhi Yuan}
\email{c.z.yuan@uestc.edu.cn}
\affiliation{Institute of Fundamental and Frontier Sciences, University of Electronic Science and Technology of China, Chengdu 610054, China}
\author{Guang-Can Guo}
\affiliation{Institute of Fundamental and Frontier Sciences, University of Electronic Science and Technology of China, Chengdu 610054, China}
\affiliation{CAS Key Laboratory of Quantum Information, University of Science and Technology of China, Hefei 230026, China}
\author{Qiang Zhou}
\email{zhouqiang@uestc.edu.cn}
\affiliation{Institute of Fundamental and Frontier Sciences, University of Electronic Science and Technology of China, Chengdu 610054, China}
\affiliation{CAS Key Laboratory of Quantum Information, University of Science and Technology of China, Hefei 230026, China}
\date{\today}
\begin{abstract}
The coexistence of quantum and classical light in the same fiber link is extremely desired in developing quantum communication. It has been implemented for different quantum information tasks, such as classical light coexisting with polarization-entangled photons at telecom O-band, and with quantum signal based quantum key distribution (QKD). In this work, we demonstrate the coexistence of energy-time entanglement based QKD and fiber optical communication at the telecom C-band. The property of noise from the classical channel is characterized with classical light at different wavelengths. With the largest noise, i.e., the worst case, the properties of energy-time entanglement are measured at different fiber optical communication rates. By measuring the two-photon interference of energy-time entanglement, our results show that a visibility of 82.01$\pm$1.10\% is achieved with a bidirectional 20 Gbps fiber optical communication over 40 km. Furthermore, by performing the BBM92 protocol for QKD, a secret key rate of 245 bits per second could be generated with a quantum bit error rate of 8.88\% with the coexisted energy-time entanglement.~Our demonstration paves the way for developing the infrastructure for quantum networks compatible with fiber optical communication.
\end{abstract} 

\maketitle
Quantum communication, which holds the promise of theoretically secure communication, is drawing more and more attention\cite{gisin2002quantum, kimble2008quantum, wengerowsky2018entanglement, wehner2018quantum, wei2022towards}. Recently, quantum communication has rapidly developed based on fiber optical communication infrastructures, where quantum states are distributed via quantum channels and the information about the measured states is exchanged via classical channels\cite{gisin2002quantum, yin2016measurement, valivarthi2019measurement}.~Most of these demonstrations have been performed with dark fiber links, in which the quantum and classical channels are deployed separately to protect the fragile quantum signals - single-photon power level from the much stronger classical signals. However, this is not a viable option under stringent operational expenditures or in fiber-scarce conditions.~Besides, the use of individual fiber links for quantum and classical optical communication would be extremely impractical for scalable quantum networks.

\begin{figure*}
    \includegraphics[width=170mm]{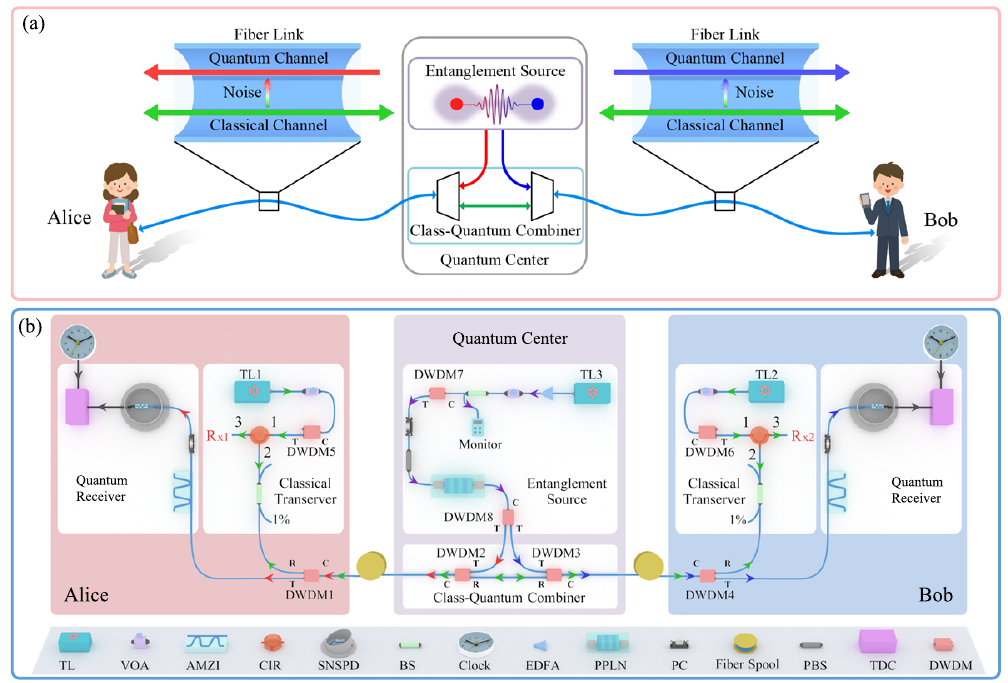}
    \caption{Scheme of the energy-time entanglement coexisting with fiber optical communication at telecom C-band. (a) Conceptual illustration of coexistence of entanglement distribution with classical optical communication. Quantum Center, consisting of the entanglement source and the DWDMs, is employed to generate entangled photons and distribute them to Alice and Bob. (b) Experimental setup for the coexistence of energy-time entangled photon distribution and classical optical communication in 40 km fiber. TL: Tunable laser, VOA: variable optical attenuator, AMZI: asymmetrical Mach-Zehnder interferometer, CIR: circulator, BS: beam splitter, CLK: clock, EDFA: erbium-doped fiber amplifier, PPLN: periodically-poled lithium niobate, PC: polarization controller, PBS: polarization beam splitter, SNSPD: superconductor nanowire single photon detector, TDC: time to digital converter, DWDM: dense wavelength division multiplexing. Note that each of the DWDM1~DWDM7 has three ports (C: common, T: transmit, R: reflect), in which the DWDM1 and DWDM2 (DWDM3 and DWDM4) transmit the photons in ITU channel C35 (C57) with a bandwidth of 100 GHz. DWDM8 has four ports, in which the two T ports transmit photons in C35 and C57, respectively.}
    \label{fig:Fig1}
\end{figure*}
Towards this end, P. D. Townsend puts forward the scheme of simultaneous transmission of quantum key distribution (QKD) and conventional data over installed fiber using wavelength-division multiplexing (WDM) for the first time in 1997\cite{townsend1997simultaneous}. A series of investigations of the coexistence of quantum signals and classical light in the same piece of fiber have been implemented\cite{tittel1998violation, chapuran2009optical, choi2010quantum, eraerds2010quantum, patel2012coexistence, patel2014quantum, wang2015experimental, kumar2015coexistence, valivarthi2016quantum, valivarthi2019measurement, berrevoets2022deployed}. Considering the extreme contrast between quantum signals and classical signals, the coexistence of them over the same fiber is always challenging. With the well-established WDM techniques, the classical signals can be readily filtered\cite{townsend1997simultaneous, choi2010quantum, patel2014quantum}. Whereas, noise photons from the coexisting classical light by means of the Rayleigh, Brillouin, or Raman scatterings and parametric optical nonlinear interactions in the fiber, mask the quantum data due to that their spectral overlap with the ones of quantum signals. One approach has been proposed to minimize such noise by locating the quantum signals at 1310 nm and the classical signals far away at 1550 nm\cite{chapuran2009optical, wang2017long, thomas2022entanglement, chung2022design, thomas2023designing}, in which quantum signals are far detuned on the anti-Stokes sideband of the classical signals. However, this happens at the expense of a higher loss and leads to a trade-off between the noise and the rate. Many studies of the coexistence of quantum signals and classical signals are proposed in which the quantum ones are also at telecom C-band using attenuated laser light\cite{burenkov2023synchronization, wang2023time}. While most quantum networks, such as entanglement-based QKD\cite{wengerowsky2018entanglement, joshi2020trusted, wen2022realizing}, quantum teleportation\cite{shen2023hertz}, quantum repeaters\cite{wei2022storage}, etc., require the distribution of quantum entanglement, which has recently been shown as a very effective resource and is naturally robust to background noise thanks to its property of quantum correlation. For instance, by tightly filtering the quantum signals in time and frequency domains, polarization-entangled quantum light can propagate coexisting with classical light over 45 km of installed fiber\cite{wu2021illinois, thomas2022entanglement}. Compared with polarization entanglement, energy-time entanglement has an innate robustness to the polarization decoherence in optical fibers, which is widely used in quantum networks\cite{cuevas2013long, lefebvre2021compact, wen2022realizing, yuan2019quantum}. However, the distribution of energy-time entanglement coexisting with classical fiber optical communication has yet been fully investigated. 

In this Letter, we investigate a coexistence scheme of energy-time entanglement distribution and classical fiber optical communication using WDM at telecom C-band. In the experiments, we measure the influence of the Raman noise from the classical channel on the quantum entanglement distribution. Our results show that the quantum correlation of photon pairs can improve the robustness of quantum channels. The energy-time entanglement distributed with a bidirectional 20 Gbps data over 40 km of optical fiber is achieved with a visibility of greater than 82.01$\pm$1.10\%. Furthermore, properties of entanglement-based QKD are analyzed with BBM92 protocol, which gives a secret key rate (SKR) of 245 bits per second (bps) with a quantum bit error rate (QBER) of 8.88\%. Our demonstration shows the potential for the development of an entanglement-based quantum network coexisting with classical fiber optical communication by employing the same infrastructure.
\begin{figure*}
    \includegraphics[width=170mm]{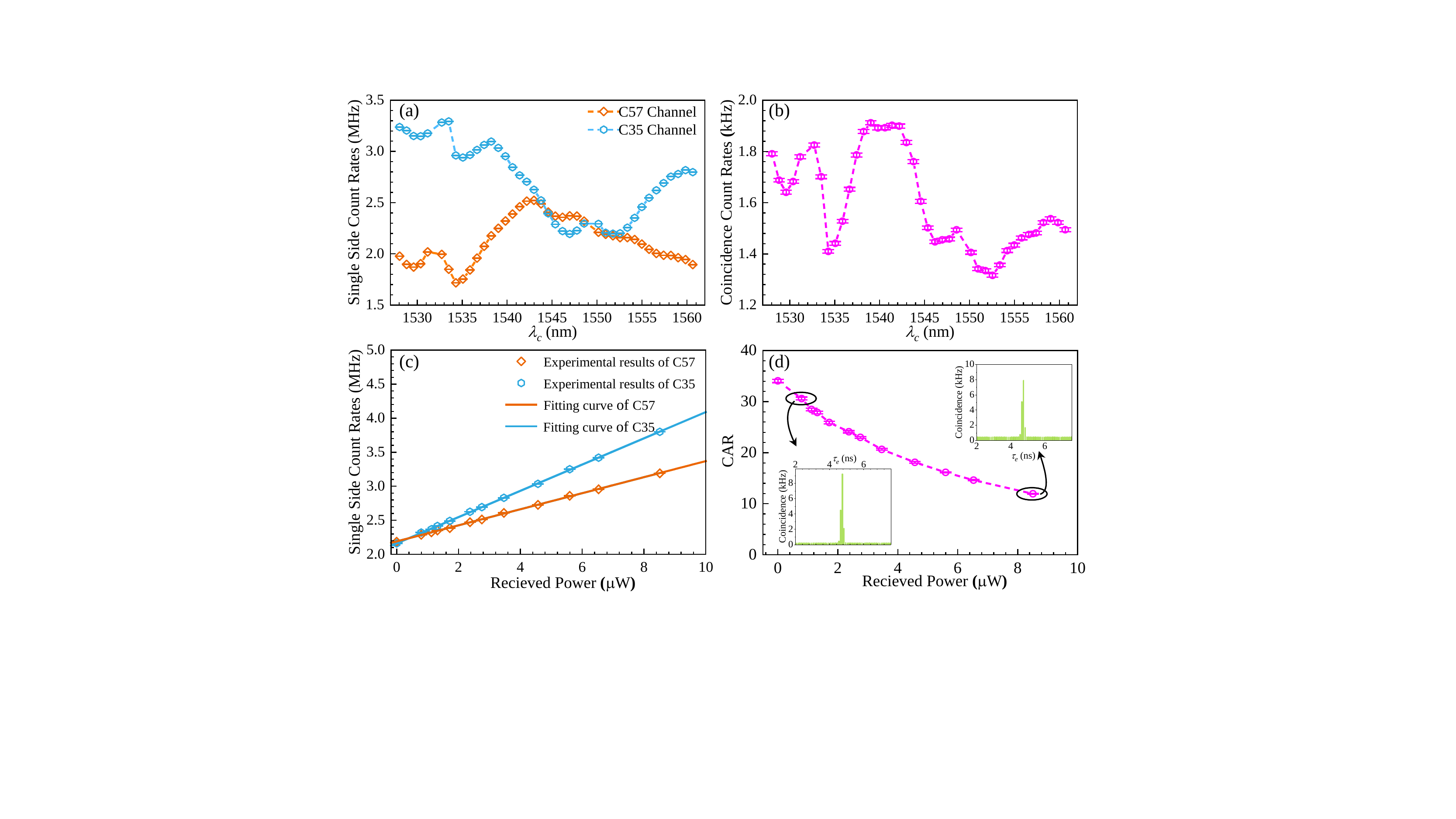}
    \caption{Experimental results of the measured Raman noises in quantum channels. (a) Single-side count rates of noise photons at C35 and C57 channels with unidirectional classical signals at different wavelengths; (b) Coincidence count rates of noise photons between C35 and C57 channels with unidirectional classical signals at different wavelengths; (c) Single-side count rates of quantum signals with different power of the classical signal at 1538.98 nm; (d) CAR with different power of the classical signal at 1538.98 nm.}
    \label{fig:Fig2}
\end{figure*}

A conceptual illustration of the coexistence of entanglement distribution with classical fiber optical communication is shown in Fig.~{\ref{fig:Fig1}}(a). Alice and Bob send and receive classical signals - as classical transceivers - through the classical channel, meanwhile, they also receive quantum signals - as quantum receivers - from the quantum center. Thus, Alice and Bob are connected through the same fiber link, in which the classical and quantum signals transmit along the same piece of fiber. The classical channels and quantum channels are located at different wavelengths. Two WDM devices (DWDM2, DWDM3) - as class-quantum combiners - are used to multiplex the classical and quantum signals in the quantum center. Although this arrangement could prevent the quantum channel from contamination of the classical channel, the Raman noise photons generated by classical signals within the quantum channels cannot be rejected. Fortunately, entangled photons exhibit inherent robustness to background noise owing to properties of correlation. By employing spectral and temporal filters and coincidence detection measurement, it is feasible to extract quantum signals from uncorrelated noise photons, which means that the quantum correlation gives immunity to the noise from classical channels.
\begin{table*}
\caption{Performances of quantum key distribution at different classical data rates.}
\begin{ruledtabular}
\begin{tabular}{cccccc}
\makecell{Classical Data} Rate & \makecell{Visibility\\$\beta=-1.571$} & \makecell{Visibility\\$\beta=-2.704$}  & Raw Key Rate&  \makecell{QBER}  & Secret Key Rate\\
\hline
0 Gbps&	88.51$\pm$1.71\% & 89.31$\pm$1.24\% &4668 bps &5.55\%& 1493 bps\\
5 Gbps&	87.28$\pm$1.21\% & 87.60$\pm$2.12\% &4712 bps &6.28\%& 1203 bps\\
10 Gbps& 85.33$\pm$0.83\%& 84.85$\pm$2.07\% &4831 bps &7.46\%& 763 bps\\
20 Gbps& 82.01$\pm$1.10\%& 82.49$\pm$2.20\% &5004 bps &8.88\%& 245 bps\\
\end{tabular}
\end{ruledtabular}
\label{tab:table2}
\end{table*}

The experimental setup of our demonstration is given in Fig.~{\ref{fig:Fig1}}(b). First, the classical light from a tunable laser at Alice (Bob), i.e., TL1 (TL2) at a wavelength of $\lambda_c$, is sent to the fiber link after propagating through a variable optical attenuator (VOA), a dense wavelength divided multiplexing device (DWDM5 or DWDM6), a circulator (Cir), and a beam splitter (BS), where the DWDMs are used to suppress the background noise from the TL1 and TL2, the Cirs are used to isolate the bidirectional classical signals, and the BSs are used for power monitoring, respectively. Second, the entangled photon pairs are generated through the cascaded second harmonic generation (SHG) and spontaneously parametric down-conversion (SPDC) processes in a single piece of periodically-poled lithium niobate (PPLN) waveguide\cite{zhang2021high} that is pumped by the TL3 with a wavelength of $\lambda_l$=1540.56 nm, i.e., ITU channel of C46. By using DWDM8, the wavelengths of signal and idler photons are selected as $\lambda_s$=1549.32 nm and $\lambda_i$=1531.90 nm, i.e., the C57 and C35 channels. Then the entangled photons and classical light are multiplexed into the same fiber links via DWDM2 and DWDM3, respectively.~After transmitting through a 20-km-long fiber link and DWDM1 (DWDM4), the signal (idler) photons enter an asymmetrical Mach-Zehnder interferometer (AMZI) to measure the property of energy-time entanglement. The photons from the AMZI are detected and recorded by superconductor nanowire single-photon detectors (SNSPDs) and a time-to-digital converter (TDC). The coexisting classical signal from Alice (Bob) to Bob (Alice) is added to a 20-km-long fiber link via DWDM1 (DWDM4), then travels to another 20-km-long fiber link via DWDM2 and DWDM3 (DWDM3 and DWDM2), and outputs from the transmission (T) port of DWDM4 (DWDM1). Finally, the power of the received classical signal is measured at the Rx2 (Rx1) as shown in Fig.~{\ref{fig:Fig1}}(b).

\begin{figure}
    \includegraphics[width=85mm]{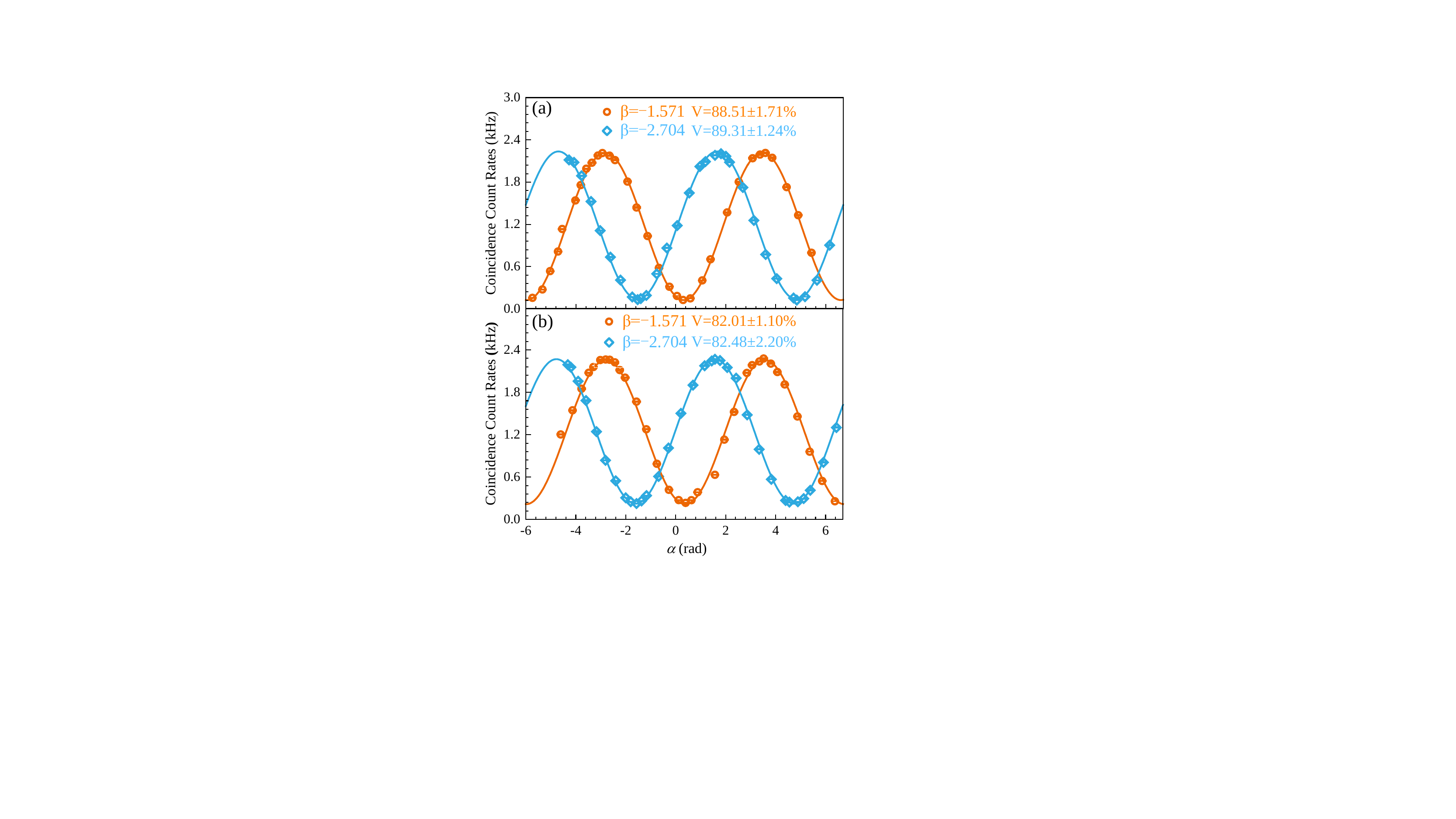}
    \caption{Quantum entanglement distribution coexisting with classical optical communication. (a) Franson interference curves of energy-time entanglement when the classical optical communication is off; (b) Franson interference curves coexisting with 20 Gbps classical optical communication.}
    \label{fig:Fig3}
\end{figure}

To investigate the property of noise from the classical channel, we first measure the single-side count rates of the Raman noise photons at C35 and C57 with unidirectional classical signals at different wavelengths.~In this case, the quantum entanglement source is disconnected at the quantum center.~When the classical signal is unidirectional from Alice to Bob, the results are shown in Fig.~{\ref{fig:Fig2}}(a), i.e., blue and orange lines, respectively.~The power of the laser is set as the receiver obtains a power of -24 dBm, i.e., the minimum required power of the 20 Gbps classical signal.~Figure~{\ref{fig:Fig2}(b) shows the coincidence count rates of noise photons between C35 and C57 channels with the classical signal at different wavelengths. 

From Fig.~{\ref{fig:Fig2}(b), the maximum coincidence count rates are observed with classical light at the wavelength of 1538.98 nm. Under this condition, i.e., the largest noise in our measurement, we measure the properties of distributed energy-time entanglement under different power, i.e., different classical data rates, of the classical signal at 1538.98 nm. The measured single-side count rates of quantum signals are shown in Fig.~{\ref{fig:Fig2}}(c). The single-side count rates of both signal and idler photons are 2.2 MHz without any classical light. With the increase of classical data rates, the single-side count rates increase linearly, which is consistence with the increase of Raman noise photon. Figure~{\ref{fig:Fig2}}(d) shows the decrease of the coincidence to accidental coincidence rate (CAR) with the increase of classical date rates. The inset of Fig.~{\ref{fig:Fig2}}(d) shows the histogram of the coincidence counts with the received classical powers of -20 and -31 dBm, respectively.

We measure the entanglement property of energy-time entangled photon pairs after being distributed from the quantum center to Alice and Bob in our experiment. Without any classical light, the Franson interference curve is shown in Fig.~{\ref{fig:Fig3}}(a). Interference visibilities of 89.31 $\pm$ 1.24\% and 88.51 $\pm$ 1.71\% are obtained with $\beta=-1.571$ and $\beta=-2.704$, respectively, where $\beta$ is the phase of the AMZI at Bob. With a classical data rate of 20 Gbps, the results of the Franson interference are shown in Fig.~{\ref{fig:Fig3}}(b) with visibilities of 82.01$\pm$1.10\% and 82.49$\pm$2.20\%, respectively. The visibilities of the Franson interference curves at different classical data rates are measured and summarized in Table~{\ref{tab:table2}}. Furthermore, we analyze the property of energy-time entanglement based QKD coexisting with classical optical communication using BBM92 protocol \cite{wen2022realizing}. In our experiment, with a classical data rate of 20 Gbps, a raw key rate of 5004 bps is obtained. The SKR is calculated as 245 bps with a QBER is 8.88\%. Performances of QKD at different classical data rates are summarized in Table~{\ref{tab:table2}}.

In conclusion, we have demonstrated the energy-time entanglement distribution coexisting with fiber optical communication at the telecom C-band. Enabled by the quantum correlation property of entangled photons, energy-time entanglement distribution coexisting with a classical data rate of 20 Gbps is achieved over 40 km of optical fiber, in which typical visibility of the Franson interference is 82.01$\pm$1.10\% after the distribution. We have achieved a QBER of 8.88\% with a SKR of 245 bps with BBM92 protocol. Our results pave the way for developing the infrastructure for quantum networks compatible with fiber optical communication.

\textit{Acknowledgments.}~This work was supported by the National Key Research and Development Program of China (Nos.~2018YFA0307400,~2018YFA0306102);~Sichuan Science and Technology Program (Nos. 2021YFSY0063,~2021YFSY0062,~2021YFSY0064, 2021YFSY0065,~2021YFSY0066,~2022YFSY0061, 2022YFSY0062,~2022YFSY0063);~National Natural Science Foundation of China (Nos. U19A2076,~62005039);~Innovation Program for Quantum Science and Technology (No. 2021ZD0301702).

\end{document}